\documentclass[aps,prl,twocolumn,twoside,floatfix,amsmath,showpacs,superscriptaddress]{revtex4-1}

\usepackage{graphicx}
\usepackage{natbib}
\usepackage{color}
\usepackage[normalem]{ulem}

\begin{document}

\newcommand{\cerusi}{CeRu$_2$Si$_2$}
\newcommand{\ybrhsi}{YbRh$_2$Si$_2$}

\newcommand{\sovert}[1]{\ensuremath{#1\,\mu\textnormal{V K}^{-2}}}
\newcommand{\resist}[1]{\ensuremath{#1\,\mu\Omega\textnormal{cm}}}
\newcommand{\mJmolK}[1]{\ensuremath{#1\,\textnormal{mJ mol}^{-1} \textnormal{K}^{-2}}}
\newcommand{\lorenzunits}{\ensuremath{\,\textnormal{W \Omega K}^{-2}}}
\newcommand{\gfac}{\ensuremath{g_{\textnormal{eff}}}}

\newcommand{\ie}{{\em i.e.}}
\newcommand{\eg}{{\em e.g.}}

\newcommand{\tred}[1]{\textcolor{red}{#1}}
\newcommand{\replace}[2]{\sout{#1} \textcolor{red}{#2}}

\title{Interplay between Kondo Suppression and Lifshitz Transitions\\in YbRh$_2$Si$_2$ at High magnetic Fields}

\author{H.~Pfau}
\affiliation{Max Planck Institute for Chemical Physics of Solids, D-01187 Dresden, Germany}
\author{R.~Daou}
\affiliation{Max Planck Institute for Chemical Physics of Solids, D-01187 Dresden, Germany}
\affiliation{Laboratoire CRISMAT, UMR 6508 du CNRS, ENSICAEN et
Universit\'{e} ́de Caen, F-14050 Caen, France.}
\author{S.~Lausberg}
\author{H.~R.~Naren}
\author{M.~Brando}
\affiliation{Max Planck Institute for Chemical Physics of Solids, D-01187 Dresden, Germany}
\author{S.~Friedemann}
\affiliation{Max Planck Institute for Chemical Physics of Solids, D-01187 Dresden, Germany}
\affiliation{Cavendish Laboratory, University of Cambridge, Cambridge CB3 OHE, United Kingdom}
\author{S.~Wirth}
\author{T.~Westerkamp}
\author{U.~Stockert}
\author{P.~Gegenwart}
\altaffiliation{I. Physikalisches Institut, Georg-August-Universit\"{a}t G\"{o}ttingen, D-37077 G\"{o}ttingen, Germany}
\author{C.~Krellner}
\altaffiliation{Physikalisches Institut, Johann Wolfgang Goethe-Universit\"{a}t, D-60438 Frankfurt am Main, Germany}
\author{C.~Geibel}
\affiliation{Max Planck Institute for Chemical Physics of Solids, D-01187 Dresden, Germany}
\author{G.~Zwicknagl}
\affiliation{Institut f\"{u}r Mathematische Physik, Technische Universit\"{a}t Braunschweig, D-38106 Braunschweig, Germany}
\author{F.~Steglich}
\affiliation{Max Planck Institute for Chemical Physics of Solids, D-01187 Dresden, Germany}

\date{\today}


\begin{abstract}

We investigate the magnetic field dependent thermopower, thermal conductivity, resistivity and Hall effect in the heavy fermion metal \ybrhsi. In contrast to reports on thermodynamic measurements, we find in total three transitions at high fields, rather than a single one at 10\,T. Using the Mott formula together with renormalized band calculations, we identify Lifshitz transitions as their origin. The predictions of the calculations show that all experimental results rely on an interplay of a smooth suppression of the Kondo effect and the spin splitting of the flat hybridized bands.

\end{abstract}

\maketitle

Since their discovery four decades ago, Kondo lattice systems pose a challenge to condensed-matter physicists. Above the characteristic Kondo temperature $T_K$ they contain a periodic lattice of local 4$f$-derived paramagnetic moments. Below $T_K$, these moments become reduced and eventually fully screened by the conduction electrons: The entanglement of the localized 4$f$ states with the delocalized conduction-band states leads to the formation of local Kondo singlets, which develop weak dispersion as a consequence of their periodic arrangement (Bloch's theorem) \cite{hewson_kondo}. The delocalized Kondo singlets act as (composite) charge carriers, which exhibit a large effective mass (heavy fermions) due to the extremely strong on-site Coulomb correlations. This is inferred from a huge Sommerfeld coefficient of the electronic specific heat \cite{trovarelli_2000} but is invisible, \eg, in photoelectron spectroscopy \cite{kummer_2011}, probing one-electron properties.

Recently, the effect of a magnetic field on these composite fermions has become an important issue. First, most information on these quasiparticles comes from de Haas-van Alphen (dHvA) experiments performed at high fields. However, band structure calculations, necessary to analyze these experimental results, exist almost exclusively at zero field \cite{miyke_2006,rourke_2008}. Second, many heavy fermion systems undergo a metamagnetic transition at fields of the order of 10\,T \cite{haen_1987,deppe_2012,schroeder_1992,gegenwart_2006}, where in some cases direct evidence (from dHvA) for a change of the Fermi surface (FS) and thus, of the heavy quasiparticles has been found \cite{aoki_1993,rourke_2008}, the origin of which remains controversial. 

\ybrhsi~is such a case: Specific heat, susceptibility and magnetostriction measurements revealed anomalies at a critical field $B_0 = 10$\,T, indicating a rapid decrease of the effective mass at $B_0$ \cite{tokiwa_2004,tokiwa_2005,gegenwart_2006}. This was initially interpreted as a breakdown of the Kondo screening. Later, dHvA experiments were interpreted in terms of a Lifshitz transition at $B_0$, where a spin-split band disappears \cite{rourke_2008,lifshitz_1960,bercx_2012}. While these two interpretations seem quite different at first sight, they essentially rely on a similar model: Only in the presence of the very flat bands of the composite fermions, the Zeeman splitting can induce such large effects on the FS at moderate fields as observed in \ybrhsi. Simultaneously a splitting also leads to a continuous decrease of the quasiparticle mass. Therefore, the field scale for both processes is related to $T_K$. 

In order to get more insight into the field evolution of the quasiparticles and into the connection between destruction of the composite fermions and the occurrence of Lifshitz transitions, we performed detailed, field dependent transport studies in \ybrhsi~in fields to at least 12\,T. We focus on the effects far above the quantum critical point (QCP) at $B = 0.06$\,T. In contrast to reports on thermodynamic measurements \cite{tokiwa_2004,tokiwa_2005,gegenwart_2006}, we show that the transition at $B_0$ consists of two close-lying features. Moreover, we observe an additional transition at 3.4\,T. From the remarkable agreement between thermopower, electrical conductivity and magnetostriction above 2\,T within the framework of Mott's formula, we demonstrate that all three transitions are caused by band structure effects.

We compare our experimental results with predictions of renormalized band (RB) structure calculations. They reveal three successive Lifshitz transitions at the three experimentally observed transition fields. This agreement demonstrates that the observed field evolution results from an interplay of a smooth suppression of the Kondo effect and the spin splitting of a sharply structured density of states (DOS) generated by a strong anisotropic hybridization. 

\begin{figure}[bht]
  \begin{center}
    \includegraphics[width=1\linewidth]{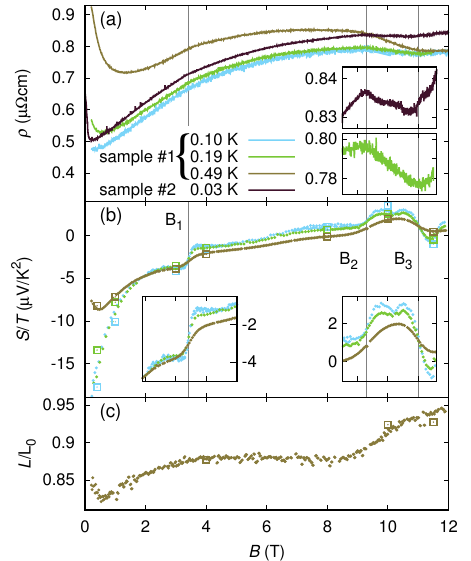}
  \end{center}
  \caption{Magnetoresistance $\rho$, thermopower $S/T$, and Lorenz ratio, $L/\mathrm{L_0}$ plotted as a function of magnetic field $B$. Lines and dots denote field sweeps; open squares are extracted from temperature sweeps. (a) $\rho$ shows three kinks at $B_1=3.4$\,T, $B_2=9.3$\,T, and $B_3=11.0$\,T. Insets: enlarged view for 0.03 and 0.19\,K. (b) $S/T$ of sample \#1 at the same temperatures as in (a). The complex behavior with several zero crossings reflects the multiband character of \ybrhsi. All three transitions are characterized by pronounced steps (insets: enlarged view). (c) The Lorenz ratio at 0.49\,K is below unity with a minimum at about 0.5\,T. }
  \label{fig:1}
\end{figure}

We investigated three high quality samples from the same batch, with residual resistivities of approximately \resist{0.5}. On sample \#1 we performed simultaneous dc~resistivity, thermal conductivity and thermopower measurements down to 0.1\,K and for 0.2\,T $\leq B \leq$ 12\,T using soldered contacts as in Ref.~\onlinecite{pfau_2012}. Sample \#2 was used to extend the resistivity data down to 0.03\,K on another setup applying an ac~technique. We also performed Hall-effect measurements on sample \#3 down to 0.05\,K and up to 15\,T. The magnetic field $B$ was always applied perpendicular to the $c$ axis. The currents for resistivity and thermal transport were parallel to $B$, for the Hall effect perpendicular to $c$ and $B$.

Figures \ref{fig:1}(a) and \ref{fig:1}(b) show the low-temperature magnetoresistivity $\rho(B)$ and the isothermal, field-dependent thermopower $S(B)$, respectively. The Lorenz ratio $L(B)/\mathrm{L_0} = \kappa(B)\rho(B) /T\mathrm{L_0}$ obtained from magnetoresistance and thermal conductivity $\kappa$ using Sommerfeld's constant $\mathrm{L}_0 = \pi^2k_B^2/3e^2$, is shown in Fig.~\ref{fig:1}(c) only for 490\,mK, because of an enhanced noise level at lower temperatures. Considering the $B$-$T$ phase diagram of \ybrhsi~\cite{gegenwart_2006}, it is natural to relate the low-field ($B<2$\,T) behavior of all these quantities to the signatures of the QCP and the surrounding non-Fermi liquid regime. These signatures are in agreement with previously reported results: the step in magnetoresistance \cite{gegenwart_2007}, the pronounced minimum in $S(B)/T$ \cite{hartmann_2010} and the minimum of the isothermal Lorenz ratio \cite{pfau_2012b} are clearly visible.

Next, we focus on the high-field properties beyond 2\,T, where \ybrhsi~is a Fermi liquid below 200\,mK \cite{gegenwart_2006}. The key features are three transitions visible as tiny kinks in the magnetoresistance at $B_1=(3.4\pm0.1)$\,T, $B_2=(9.3\pm0.1)$\,T and $B_3=(11.0\pm0.2)$\,T in Fig.~\ref{fig:1}(a). Their position is sample and temperature independent, cf.~also \cite{pourret_2013}. The transitions are more obvious in the thermopower (Fig.~\ref{fig:1}(b)): $S(B)/T$ shows three pronounced steps, which become sharper at lower $T$. 

These observations are interesting from two perspectives. First, previous measurements have not observed features at 3.4\,T \cite{tokiwa_2004,tokiwa_2005,gegenwart_2006}. Second, the transition reported in magnetization, specific heat and magnetostriction \cite{tokiwa_2004,tokiwa_2005,gegenwart_2006} at roughly 10\,T is actually composed of two well-separated features at 9.3\,T and 11.0\,T, which do not merge in the extrapolation $T\rightarrow 0$.

We further analyze our data using the Mott formula \cite{mott_1971} for the diffusion thermopower to clarify if the observed features have a thermodynamic origin. Phonon drag contributions to the thermopower are negligible in this temperature range \cite{hartmann_diss}. The Mott formula generally holds at low temperatures in the absence of inelastic scattering \cite{johnson_1980}. To expand it, we first exchange the energy derivative of $\ln{\sigma}$ with the field derivative using $\partial B/ \partial \epsilon$. In the second step, we use $\sigma = ne^2\tau/m^*$ together with $m^*\propto N^{2/3}$ (in 3D) and obtain

\begin{equation}
 \frac{S}{e\mathrm{L_0}T}
   = \frac{\partial B}{\partial\epsilon}\frac{\partial \ln{\sigma}}{\partial B}\biggr |_{\epsilon_\mathrm{F}}
 = \frac{\partial B}{\partial\epsilon}
\left(\frac{\partial \ln{\tau}}{\partial B} - \frac{2}{3}\frac{\partial \ln{N}}{\partial B}  \right)_{\epsilon_\mathrm{F}}
 \label{eqn:mott-field}
\end{equation}
where $\sigma = 1/\rho$ is the electrical conductivity, $n$ the total electron concentration, $N$ the DOS, $m^*$ the effective mass, and $\tau$ the scattering time. This splits the thermopower into a scattering part ($\tau$) and a part representing the band structure ($N$).

\begin{figure}[tbh]
  \begin{center}
    \includegraphics[width=1\linewidth]{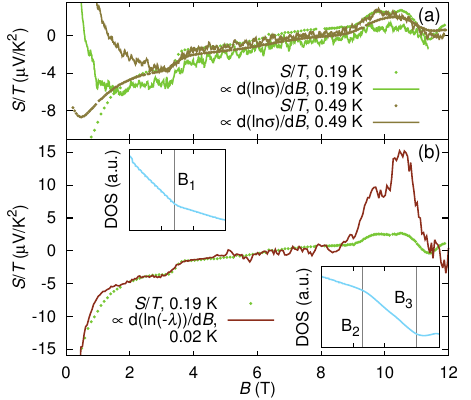}
  \end{center}
  \caption{Analysis of the thermopower with the Mott formula. The measured $S/T$ (dots) is compared with that calculated (lines) from (a) the electrical conductivity and from (b) the magnetostriction coefficient using Equation~\ref{eqn:mott-field} with $\partial\ln{N}/\partial B \propto \partial \ln{\lambda}/{\partial B}$ and $\partial\ln{\tau}/\partial B = 0$. Within the Fermi liquid regime ($B>2-3$\,T) all curves show the same overall behavior. Insets: Calculated DOS at $\epsilon_\mathrm{F}$ using thermopower data at 0.1\,K. For the left inset $+S$ and for the right inset $-S$ was integrated to match the band structure calculations from Ref.~\onlinecite{zwicknagl_2011}.}
  \label{fig:compareS}
\end{figure}

To test the first expansion in Equation \ref{eqn:mott-field}, we compare $S(B)/T$ with $\partial \ln{\sigma}/\partial B$ in Fig.~\ref{fig:compareS}(a). Above 3\,T, both curves can be scaled on top of each other using a constant $\partial B/\partial\epsilon$. This suggests a linear relationship between $\epsilon$ and $B$, which we write as an effective Zeeman energy $\epsilon = \gfac\mu_B B/2$, with $\gfac$ of $16\pm1$ ($\mu_B$ is the Bohr magneton), and discuss later on. Moreover, it implies that inelastic scattering (which would invalidate the Mott formula) is insignificant in this regime.
In the low-field region below 2\,T both curves show a disparate behavior. Consequently, either $\epsilon\propto B$ is violated or inelastic scattering is significant. Both are likely to occur close to a QCP: The first can arise if there are dramatic changes in the magnetization, the band structure or the FS. Inelastic scattering can increase close to a phase transition.

We now turn to the second expansion of Equation~\ref{eqn:mott-field}. Several thermodynamic probes give access to the contribution from the DOS, while being independent of scattering effects. We use the linear magnetostriction coefficient $\lambda$, since the data available have a higher resolution than \eg~specific heat or magnetization. Applying a constant scaling factor, $\partial \ln{\lambda}/{\partial B}$ matches $S/T$ in Fig.~\ref{fig:compareS}(b), which implies a power law $\lambda \propto N^{\alpha}$. Only the double hump around 10\,T is more pronounced. Considering the nice agreement at low fields, 
it is likely that the discrepancy of $S$ and $\sigma$ in this regime is due to enhanced inelastic scattering rather than a failure of $\epsilon \sim B$. The surprisingly good qualitative agreement, especially at $B_1$, $B_2$, and $B_3$, proves that the origin of all three transitions lies within the correlated band structure of \ybrhsi.

Our experimental data together with previous results already indicate the complex nature of these band structure effects. 
For example, the transition around $B_0$ is composed of two fields and the Sommerfeld coefficient $\gamma(B)$ decreases only moderately between them from 250\,mJ/molK$^2$ to 100\,mJ/molK$^2$. It is therefore unlikely that either a single Lifshitz transition \cite{rourke_2008} or a sudden localization of the $f$ electrons \cite{tokiwa_2004,tokiwa_2005,gegenwart_2006} is---on its own---a sufficient model to describe this behavior. Hence, a theory of the field evolution should include not only the Kondo effect to describe derenormalization processes, but also the specific correlated band structure to reveal topological transitions.

\begin{figure}[bht]
  \begin{center}
    \includegraphics[width=1\linewidth]{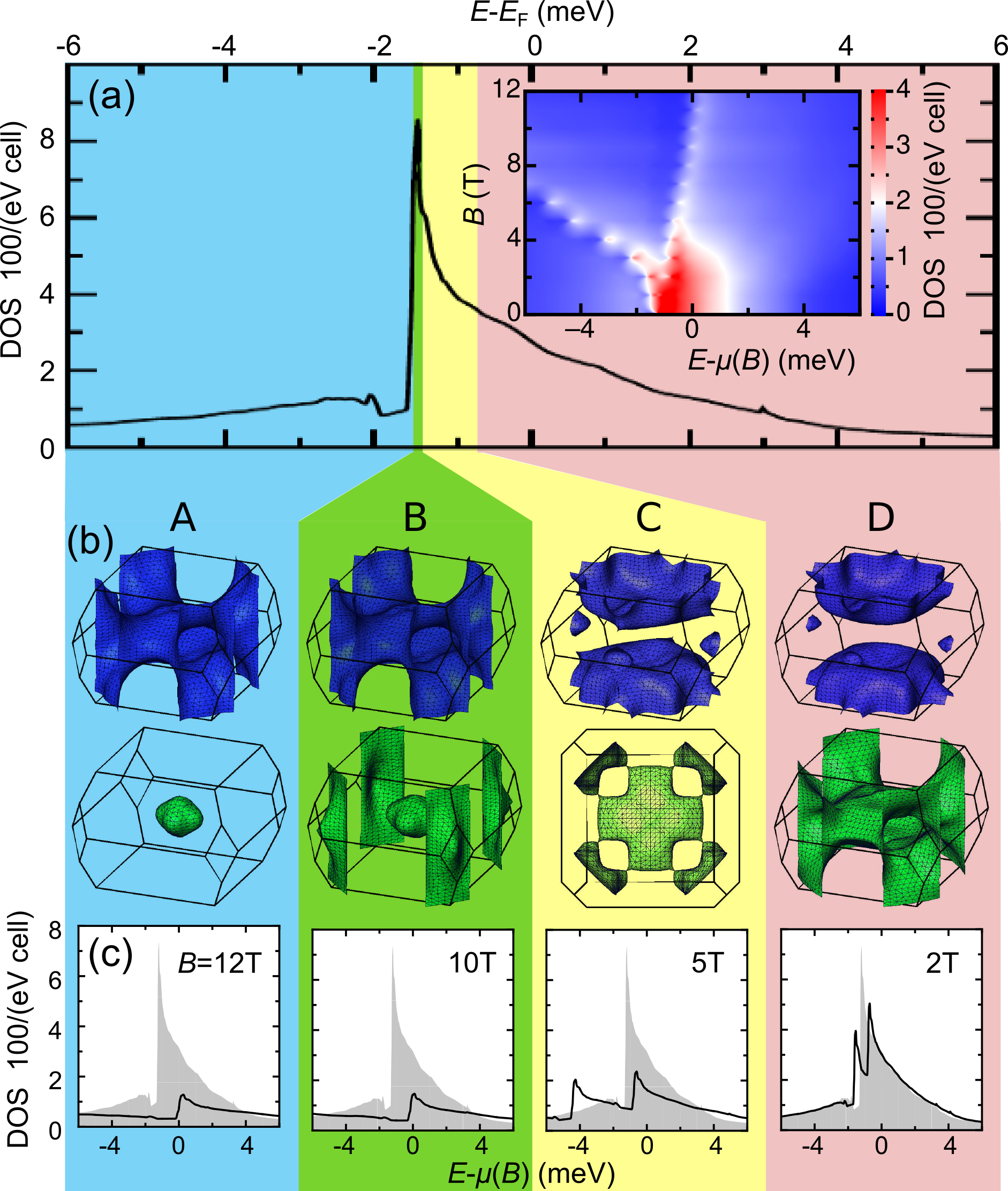}
  \end{center}
  \caption{Isoenergy surfaces for $B=0$ and the quasiparticle DOS development in finite field calculated using the RB method. The zero-field DOS in (a) is divided into four regions (blue (A), green (B), yellow (C), red (D)) distinguished by different topologies of the isoenergy surfaces shown in (b). In the yellow region (C), we show in plan the ``jungle gym'' FS sheet exactly at the topological transition between B and C. (c) illustrates the magnetic field evolution of the DOS for selected fields, with the zero-field DOS in gray for comparison. Inset in (a): Energy-field map of the DOS interpolated in 1\,T steps. One can assign the four energy regions and their isoenergy surfaces in a) and b) to the four field ranges and their FS separated by $\widetilde{B}_1$, $\widetilde{B}_2$ and $\widetilde{B}_3$.}
  \label{fig:dos_and_fs}
\end{figure}

We therefore conducted field-dependent RB calculations (Fig.~\ref{fig:dos_and_fs}), which are an extension to the results from Refs.~\onlinecite{zwicknagl_2011}, focused on the detailed development of the FS and the DOS in field. We used the RB method described in Refs.~\onlinecite{zwicknagl_2011,zwicknagl_1992} with a tight $k$ mesh in zero field of 8125 points in the irreducible wedge to resolve changes in the isoenergy surfaces. In finite fields we used 405 $k$ points. 

Figure \ref{fig:dos_and_fs}(a) shows the calculated zero-field quasiparticle DOS $N(\epsilon)$. Fig.~\ref{fig:dos_and_fs}(b) displays the corresponding isoenergy surfaces of the most important FS sheets with $f$ character---the so called ``doughnut''(top) and the ``jungle gym'' (bottom), respectively. Scanning through $N(\epsilon)$ within the displayed energy range, the calculations predict the four color-coded regimes characterized by different topologies of the isoenergy surfaces, separated by Lifshitz transitions. 

An applied magnetic field spin-splits the DOS into a minority and a majority branch. These do not shift rigidly in field but the amplitude of the coherence peak decreases (Fig.~\ref{fig:dos_and_fs}(c)), mainly because of the weakening of the Kondo effect. Nevertheless the characteristics of the band structure are not changed. Therefore we expect the energy evolution of the isoenergy surfaces in Fig.~\ref{fig:dos_and_fs}(b) to be the same as the field evolution of the FS. {\em I.e.}, the FS of the majority band stays in the red (D) regime, while the FS topology of the minority band changes from the red (D) type through yellow (C) and green (B) to the blue one(A).

We take advantage of the similar shape of the zero- and finite-field DOS and assign the energy of a topology change (in zero field) to a magnetic field where the corresponding feature in the DOS crosses $\epsilon_\mathrm{F}$: The transition from the blue (A) to the green (B) regime corresponds to a kink in the DOS which reaches $\epsilon_\mathrm{F}$ for $\widetilde{B}_3 = (11\pm 1)$\,T. Similarly, we obtain $\widetilde{B}_2 = (9\pm 1)$\,T for the second transition (green (B)---yellow (C)). The difference $\widetilde{B}_3-\widetilde{B}_2$ fits to the linear shift of 0.1\,meV/T (inset Fig.~\ref{fig:dos_and_fs}(a)), which in turn matches the ESR $g$ factor of 3.5~\cite{schaufuss_2009} (applying $\widetilde{\epsilon} = g\mu_B B/2$). We use this shift to estimate the field corresponding to the weak third transition from yellow (C) to red (D) to $\widetilde{B}_1 = (2.5\pm 1)$\,T. These fields, extracted from the calculations, are in excellent agreement with the transitions $B_1$, $B_2$ and $B_3$ found 
experimentally. This proves that they correspond to three Lifshitz transitions of the types illustrated in Fig.~\ref{fig:dos_and_fs}(b).

Additionally, the linear sweep of the van Hove singularity predicted by our calculations confirms the linear energy-field dependence found in our data analysis. The only adjustable parameter leading to the remarkable accuracy of the Mott formula in the two comparisons shown in Fig.~\ref{fig:compareS} is $\gfac$. Moreover, the thermopower as well as $\partial\ln{\lambda}/{\partial B}$ and $\partial \ln{\sigma}/{\partial B}$ are independent of sample geometry; thus, systematic errors are almost negligible. The difference between $\gfac = 16$ (corresponding to a $\partial \epsilon/\partial B=0.5$\,meV/T) and $g=3.5$ ($0.1$\,meV/T) from our calculations, however, is not surprising, since a rigid band shift is obviously insufficient to account for the experimental data. Unexpectedly, these field-induced changes of the band structure also enter linearly into $\gfac$.

\begin{figure}
  \begin{center}
    \includegraphics[width=1\linewidth]{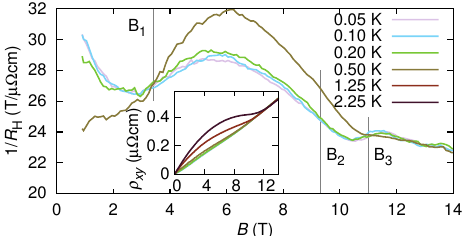}
  \end{center}
  \caption{Hall effect on sample \#3. The field dependent inverse Hall coefficient shows several extrema close to $B_1$,$B_2$ and $B_3$ obtained from resistivity data (see Fig.~\ref{fig:1}(a)). Inset: the Hall resistivity, $\rho_{xy}$, is almost linear below 0.2\,K.}
  \label{fig:hall}
\end{figure}

To include an additional link between our experimental data and the RB calculations, we draw upon the good qualitative agreement between $S/T$ and $\lambda$ in Figure \ref{fig:compareS}(b) and calculate the field dependent DOS at $\epsilon_\mathrm{F}$ straightforwardly from the thermopower measurements by integrating $S$ over $B$ (ignoring $(\partial \ln{\tau}/\partial B)$ in Equation \ref{eqn:mott-field}). Importantly, the so-obtained DOS (Figure \ref{fig:compareS}(b)) matches the features at the transition fields as calculated by the RB method (see Figure 5 in Ref. \onlinecite{zwicknagl_2011}): a kink at $B_1$ and a steplike decrease between $B_2$ and $B_3$.

Our results agree with the low-temperature heavy Fermi liquid state observed at $B>0.06$\,T, \ie~in the paramagnetic Kondo-lattice phase. While, according to Numerical Renormalization Group results for a single impurity (see \eg~\cite{hewson_2004,hewson_2006,bauer_2007,peters_2006}), the suppression of the on-site Kondo screening by a magnetic field implies a continuous decrease of the effective mass $m^*$, we observe abrupt changes in thermodynamic and transport properties related to $m^*$. Therefore, the single impurity model alone cannot account for the observations reported here. The latter have to be attributed to coherence effects arising from the periodic arrangement of the Kondo ions and are well explained by our RB calculations. The anisotropic hybridization of the $4f$ states with the conduction bands, caused by the highly anisotropic crystalline electric field (CEF) ground state, leads to van Hove-type singularities in the quasiparticle DOS. The structures in the quasiparticle DOS highlight changes of the isoenergy surfaces as shown in Fig.~\ref{fig:dos_and_fs}. In a magnetic field, the $4f$ states are split which, in turn, leads to a Zeeman splitting of the quasiparticle bands. The relative shifts of the latter, however, are enhanced by a field-dependent Sommerfeld-Wilson ratio which reflects the local many-body effects.

Experimentally, further insight into the evolution of the FS can be obtained by Hall effect measurements.
The isothermal Hall resistivity $\rho_{xy}(B)$ for selected $T$ is presented in the inset of Fig.~\ref{fig:hall} and indicates hole-dominated transport. Importantly, beyond 12 T all curves coincide pointing towards a field-driven suppression of the local Kondo effect. This is a continuous process and likely accounts for the maximum in the effective carrier concentration, $1/R_\mathrm{H} = 1 / \frac{d \varrho_{xy}}{dB}$, at around 6 T and lowest temperatures via the hybridization of conduction and 4$f$ electrons. For $T>0.2$\,K an anomalous contribution to $R_\mathrm{H}$ comes into play (\cite{coleman_1985, paschen}) such that $1/R_H$ no longer tracks the carrier concentration. The transitions at $B_1$, $B_2$ and $B_3$ appear also in the transverse magnetoresistance ($j \perp B$, not shown) measured on sample \#3 with the same signatures as in the longitudinal magnetoresistance (Fig.~\ref{fig:1}(a)). However, the extrema in $R_H(B)$ simultaneously measured on sample \#3  are at slightly different fields. The latter might be caused by the complex nature of the Hall effect \cite{nair_2012} ({\it e.g.} multiple bands).

In conclusion, we showed that the thermopower is particularly suitable for revealing field induced changes in the FS of a correlated metal, hardly detectable by any other probe. In \ybrhsi, we find three successive transitions, which were identified as Lifshitz transitions by a comparison with predictions from detailed RB calculations. This implies, that the unusual high-field properties of \ybrhsi~arise from the interplay of (a) the symmetry of the CEF ground state ($g$ factor, anisotropic hybridization), (b) the suppression of the local Kondo effect (reduced effective mass, field-dependent Sommerfeld-Wilson ratio) and (c) the coherence effects due to the periodicity of the lattice (van Hove singularities, Lifshitz transitions). The excellent agreement between our experimental results and our RB calculations demonstrates that RB calculations are a very suitable approach to describe quasiparticles in the Kondo lattice.

We acknowledge insightful discussions with F.~F.~Assaad, M.~Bercx, M.~Garst and P.~Thalmeier. Part of this work has been supported by the DFG Research Unit 960 “Quantum Phase Transitions.”

\bibliographystyle{unsrt}

\end{document}